\documentclass[12pt]{article}
\usepackage{amsmath,amssymb,graphicx,slashbox}

\def\mydate{March 11, 2009}
\def\ignore#1{{}}

\ignore{
\setlength{\textwidth}{150mm} 
\setlength{\textheight}{230mm} 
\setlength{\oddsidemargin}{5mm} 
\setlength{\evensidemargin}{5mm} 
\setlength{\topmargin}{-5mm} 
\makeatletter
  
  \@addtoreset{equation}{section}
\makeatother

}

\renewcommand{\baselinestretch}{1.25}

\renewcommand{\thefootnote}{\arabic{footnote}}

\newcommand{\beeq}{\begin{equation}}
\newcommand{\eneq}{\end{equation}}
\newcommand{\beqn}{\begin{eqnarray}}
\newcommand{\eeqn}{\end{eqnarray}}

\def\mybig{\displaystyle \strut }

\def\dd{\partial}
\def\la{\raise.16ex\hbox{$\langle$}\lower.16ex\hbox{}  }
\def\ra{\, \raise.16ex\hbox{$\rangle$}\lower.16ex\hbox{} }
\def\go{\rightarrow}

\def\onehalf{ \hbox{${1\over 2}$} }

\def\eff{{\rm eff}}

\def\cL{{\cal L}}

\def\EM{{\rm EM}}

\def\KK{{\rm KK}}

\def\vphi{\varphi}

\def\psibar{ \psi \kern-.65em\raise.6em\hbox{$-$} }
\def\psibarl{ \psi \kern-.65em\raise.6em\hbox{$-$} \lower.6em\hbox{} }

\def\myfrac#1#2{{\mybig #1\over \mybig #2}}

\begin{document}

\thispagestyle{empty}

\baselineskip=12pt

{\small \noindent \mydate    \hfill OU-HET 620/2008}

{\small \noindent (Corrected.)    \hfill}


\baselineskip=35pt plus 1pt minus 1pt

\vskip 2.5cm

\begin{center}
{\Large \bf Yukawa Couplings and Effective Interactions}\\
{\Large \bf in  Gauge-Higgs Unification}\\

\vspace{2.5cm}
\baselineskip=20pt plus 1pt minus 1pt

{\def\thefootnote{\fnsymbol{footnote}}
\bf 
Yutaka Hosotani  and Yoshikazu Kobayashi }\\

\vspace{.3cm}
{\small {\it Department of Physics, Osaka University,
Toyonaka, Osaka 560-0043, Japan}}\\
\end{center}

\vskip 2.0cm
\baselineskip=20pt plus 1pt minus 1pt

\begin{abstract}
The wave functions and Yukawa couplings of the top and bottom quarks 
in the $SO(5) \times U(1)$ gauge-Higgs unification model are determined.
The result is summarized in the effective interactions for 
$\hat \theta_H(x) = \theta_H + H(x)/f_H$  
where $\theta_H$ is the  Wilson line phase and $H(x)$ is the 4D Higgs field.
The Yukawa,  $WWH$ and $ZZH$ couplings vanish at $\theta_H = \onehalf \pi$.
There emerges the possibility that the Higgs particle becomes stable. 
\end{abstract}


\newpage

In the standard model of electroweak interactions the electroweak (EW) 
symmetry is spontaneously broken by the Higgs field, 
the mechanism of which is yet to be scrutinized and confirmed by experiments.  
The Higgs particle  is expected to be found at LHC in the coming years.  
It is not clear at all, however,  if the Higgs particle
appears as described in the standard model.  It is often argued from a theoretical
point of view that the naturalness and stability against radiative corrections
to the Higgs field indicate the existence of supersymmetry underlying the nature.  
Other scenarios with the naturalness  have also been proposed, 
among which are the little Higgs theory, the Higgsless model,
and the gauge-Higgs unification scenario.\cite{Csaki1, Cheng1, YHsusy} 

Recently there has been significant progress in the gauge-Higgs unification scenario
in which the 4D Higgs field is identified with a part of the extra-dimensional
component of gauge fields in higher dimensions.\cite{Fairlie1}-\cite{Hatanaka1}
The Higgs field appears as an Aharonov-Bohm (AB) phase, or a Wilson line phase,
in the extra dimension,  thereby  the EW symmetry being dynamically broken 
by the  Hosotani mechanism.\cite{YH1, YH2, Davies1}
The $SO(5)\times U(1)_X$ gauge-Higgs unification model in the Randall-Sundrum
(RS) warped space-time has been extensively studied to give definitive 
predictions.\cite{Agashe1}-\cite{HOOS}

The nature of the Higgs field as an AB phase plays a decisive role here.  
Let us denote the Wilson line phase along the extra dimension by $\theta_H$.
The effective potential  $V_\eff (\theta_H)$ becomes finite at the one loop level
thanks to the AB phase nature of $\theta_H$.
The  neutral Higgs field $H(x)$ corresponds to four-dimensional fluctuations
of $\theta_H$.  It immediately follows that the Higgs mass, related to the curvature
of $V_\eff$ at the minimum, is predicted at a finite value, once the matter
content of the theory is specified.  Another distinctive prediction is obtained
for the Higgs couplings to $W$ and $Z$.  In the RS warped spacetime
the $WWH$ and $ZZH$ couplings are suppressed by a factor $\cos \theta_H$
compared with those in the standard model.\footnote{It has been discussed 
that the suppression occurs in a wider class of models.\cite{Giudice1}}

Inclusion of quarks and leptons, particularly of top and bottom quarks, 
is crucial to have EW symmetry breaking.  Medina, Shar, and Wagner (MSW)
proposed an  $SO(5)\times U(1)_X$ gauge-Higgs unification model with 
top and bottom quarks in which the EW symmetry breaking is induced.\cite{MSW}
More recently Hosotani, Oda, Ohnuma and Sakamura (HOOS) have proposed a 
model with simpler matter content and many predictions.\cite{HOOS}
It has been shown there that $V_\eff (\theta_H)$ is minimized at 
$\theta_H = \onehalf \pi$ and the Higgs mass is predicted around 50 GeV.
The LEP2 bound for the Higgs  mass is evaded 
thanks to the vanishing $ZZH$ coupling  at $\theta_H = \onehalf \pi$.

The  purpose of the present paper is two-fold.  The Yukawa couplings
of quarks to the 4D Higgs field stem from gauge interactions in the extra-dimension.
We first evaluate the 4D Yukawa couplings in the HOOS model 
in the Kaluza-Klein approach by determining 
the wave functions of the Higgs field and quarks,  inserting them into 
the five-dimensional action, and integrating over the extra-dimensional coordinate.   
Secondly we develop an effective interaction approach for 
the Higgs couplings to quarks.  As the Higgs field is a fluctuation mode
of $\theta_H$, the Yukawa couplings are related to the $\theta_H$-dependence 
of the masses of quarks in this approach.   We shall see that
the Yukawa couplings in the HOOS model determined in these two approaches 
coincide with each other with high accuracy.    
This establishes the validity of the effective
interactions at low energies, which enables us  to deduce  higher-order Higgs 
couplings such as $H^n t \bar t$ by bypassing  laborious procedure of 
summing over contributions of intermediate Kaluza-Klein (KK) excited states.

We analyze the $SO(5) \times U(1)_X$ model with top and bottom quarks specified 
in ref.\ \cite{HOOS}, following the notation there.
The model is defined in the Randall-Sundrum  (RS) warped spacetime
whose metric is given by
\beeq
ds^2 = \frac{1}{z^2} \Big\{ \eta_{\mu\nu} dx^{\mu}dx^{\nu} +  \frac{dz^2}{k^2} \Big\}
\label{metric1}
\eneq
for $1 \le z \le z_L$.  The bulk region $1 < z < z_L$ is  an AdS spacetime with the 
cosmological constant  $\Lambda = - 6k^2$,  being sandwiched 
by the Planck brane at $z=1$ and by the TeV brane at $z=z_L$.  
The warp factor $z_L$ is large, typically around $10^{13}$ to $10^{17}$.  
The $SO(5) \times U(1)_X$ gauge symmetry is broken  to $SO(4) \times U(1)_X$
by the orbifold boundary conditions at the Planck and TeV branes with the 
parity matrices given by $P_0=P_1={\rm diag}(-1,-1,-1,-1,1)$.
The symmetry is further broken to $SU(2)_L \times U(1)_Y$ by additional
interactions at the Planck brane.  

The 4D Higgs field appears as a zero mode in the $SO(5)/SO(4)$ part 
of the fifth  dimensional component of the vector potential $A_z^{\hat a} (x,z)$
$ (a=1,\cdots,4)$,  which is expanded as 
\beeq
A^{\hat{a}}_z(x,z)=\phi^{a}(x) \vphi_H(z) +\cdots
~,~~ \vphi_H(z) = \sqrt{\myfrac{2}{k(z^2_L-1)}} \, z ~.
\label{Higgs1}
\eneq
An $SO(4)$ vector $\phi^a$ forms an $SU(2)_L$ doublet
$\Phi_H(x)^t = (1/\sqrt{2}) (\phi^2+i\phi^1 , \phi^4 -i\phi^3)$ corresponding
to the Higgs doublet in the standard model.
Without loss of generality one can assume  $\la \phi^a \ra =v\delta^{a4}$ 
when the EW symmetry is spontaneously broken by the Hosotani mechanism. 
Let us denote the generators of $SO(5)/SO(4)$ by $T^{\hat{a}}$ $ (a=1,\cdots,4)$.
In the vectorial representation
$(T^{\hat 4})_{ab} =( i /\sqrt{2}) (\delta_{a5} \delta_{b4} -\delta_{a4} \delta_{b5}) $,
whereas in the spinorial representation $T^{\hat 4} = ( 1 /2\sqrt{2}) I_2 \otimes \tau_1$.
The Wilson line phase $\theta_H$ is given by 
$\exp\{\frac{i}{2}\theta_H (2\sqrt2 \, T^{\hat{4}})\}
=\exp\{ ig_A \int^{z_L}_{1}dz \la A_z\ra \}$ so that
\begin{equation}
\theta_H =\frac12 g_A v \sqrt{\frac{z^2_L-1}{k}}
\sim\frac{g v}{2}\frac{\pi\sqrt{kL}}{m_\KK}.
\label{Higgs2}
\end{equation}
Here the $SO(5)$ gauge coupling constant $g_A$ in five dimensions is related to 
the four-dimensional $SU(2)_L$ gauge coupling constant $g$ by  
$g= g_A/\sqrt{L}$ where $L= k^{-1} \ln z_L$ is the size of the fifth dimension
in the $y$  $(\equiv  k^{-1} \ln z)$ coordinate.
The Kaluza-Klein mass scale is given by 
$m_\KK = \pi k (z_L -1)^{-1} \sim \pi k z_L^{-1}$.
The $W$ boson mass is approximately given by 
$m_W \sim \sqrt{k/L} \,  z_L^{-1} |\sin\theta_H |$.  
The value for $\theta_H$ is dynamically determined 
such that the effective potential $V_\eff (\theta_H)$
is minimized.   In the HOOS model $\theta_H = \onehalf \pi$.  
With $m_W$ and $z_L$ given,  $k$  and $m_\KK$ are fixed.
For $z_L = 10^{13}$ to $10^{17}$, $k$ ranges from
$4.4 \times 10^{15}\,$GeV to $5.0 \times 10^{19}\,$GeV,
but $m_\KK$ varies only from 1.38$\,$TeV to 1.58$\,$TeV.
Physics predictions do not sensitively depend on the parameter $z_L$
in this range.

The main focus in the present paper is given on fermions and their
interactions.  Let us consider fermion multiplets
containing top and bottom quarks. 
In the bulk region $1<z<z_L$ two $SO(5)$ vector multiplets, $\Psi_a$ 
$(a=1,2)$,  are introduced with the action
${\cal L}_{\rm bulk}^{\rm fermion} = \sum_{a=1}^2 \onehalf \big\{
\overline{\Psi}_a {\cal D}(c_a) \Psi_a +  {\rm h.c.} \big\}$
where $c_a$ denotes the dimensionless bulk mass parameter.  
Each of $\Psi_a$'s consists of $SO(4)$ vector
and singlet components.  The former is decomposed into
two $SU(2)_L$ doublets  with $SU(2)_R$ charges
$T^{3_R} = \pm \onehalf$;
\beqn
&&\hskip -1cm
\Psi_1 = \bigg[  \begin{pmatrix} T \cr B \end{pmatrix} \equiv Q_1 , 
\begin{pmatrix} \, t \,  \cr b \end{pmatrix} \equiv q ,  
\, t' \, \bigg]_{\frac{2}{3}} ~, \cr
\noalign{\kern 5pt}
&&\hskip -1cm
\Psi_2 = \bigg[ \begin{pmatrix} U \cr D \end{pmatrix} \equiv Q_2 , 
 \begin{pmatrix} X \cr Y \end{pmatrix} \equiv Q_3 ,  
\, b' \bigg]_{-\frac{1}{3}} ~.
\label{fermion1}
\eeqn
The subscript $\frac{2}{3}$ or $-\frac{1}{3}$ indicates the $U(1)_X$ charge $Q_X$.
The electric charge is given by $Q_E = T^{3_L} + T^{3_R} + Q_X$.  
The orbifold boundary condition is given by
$\Psi_a(x , y_j -y) =  P_j \Gamma^5 \Psi_a (x, y_j + y)$ in the $y$
coordinate with $(y_0 , y_1) = (0, L)$.   This leads to zero modes
in $Q_{aL}$, $q_{aL}$, $t'_R$ and $b'_R$, where the subscripts $L$ and $R$
denote the left- and right-handed components in four dimensions, respectively. 

In addition to the bulk fermions, 
three right-handed multiplets  localized on the Planck brane,
belonging to $(\onehalf, 0)$  representation of 
$SU(2)_L \times SU(2)_R$,  are introduced;
\beeq
\hat \chi_{1R} = \begin{pmatrix} \hat T_R \cr \hat B_R  \end{pmatrix}_{7/6} ,~
\hat \chi_{2R} = \begin{pmatrix} \hat U_R \cr \hat D_R  \end{pmatrix}_{1/6} ,~
\hat \chi_{3R} = \begin{pmatrix} \hat X_R \cr \hat Y_R  \end{pmatrix}_{-5/6} .
\label{fermion2}
\eneq
Here the subscripts $7/6$ etc.\ represent the $U(1)_X$ charges.
The brane fermions $\hat \chi_{aR}$ have, besides gauge invariant kinetic
terms on the Planck brane, mass terms with $q_L$ and $Q_{aL}$ given by
\beeq
\cL_{\rm mass}^{\rm brane} = - i \delta (y)   \bigg\{
 \sum_{\alpha =1}^3  \mu_\alpha 
\hat \chi_{\alpha R}^\dagger Q_{\alpha L}
+ \tilde \mu  \hat \chi_{2 R}^\dagger \, q_L  \bigg\} + ({\rm h.c.}) ~.
\label{fermion3}
\eneq
The four brane mass parameters, $\mu_\alpha$ and $\tilde \mu$
 have dimensions of (mass)$^{1/2}$.  
We suppose that $\mu_\alpha^2, \tilde \mu^2 \gg m_\KK$.
In this case   the only relevant parameter for the spectrum at low energies 
turns out the ratio $\tilde \mu/\mu_2 \sim m_b/m_t$.

In ref.\ \cite{HOOS} the spectrum of various fields were determined
in the twisted gauge  achieved by a gauge transformation
\beeq
\Omega(z) = \exp \big\{  i \theta (z) \sqrt{2} \, T^{\hat 4} \big\} ~~,~~
\theta (z) = \frac{z_L^2 - z^2}{z_L^2 - 1} \, \theta_H ~~.
\label{gaugeT1}
\eneq
In the twisted gauge $\tilde A_M = \Omega A_M \Omega^\dagger
- (i/g) \Omega \dd_M \Omega^\dagger$ and  the background field
 vanishes, $\la \tilde A_M \ra = 0$,  but 
the boundary conditions at $z=0$ get twisted from the original ones.

The fields in the bulk satisfy the free equations in the linear approximation.
The equations in the bulk for the fermion fields
 $\tilde \Psi \equiv z^{-2} \, \Omega \, \Psi$ 
with the bulk mass parameter $c$ simplify to 
\beeq
\left\{ \begin{pmatrix} & \sigma \dd \cr \bar \sigma \dd \end{pmatrix} 
- k \begin{pmatrix} D_- (c) \cr & D_+ (c) \end{pmatrix}  \right\}
\begin{pmatrix} \tilde \Psi_R \cr \tilde \Psi_L \end{pmatrix} = 0 
\label{Feq1}
\eneq
where $D_\pm (c) = \pm(d/dz)  + (c/z)$.  Various fields mix among
themselves through the brane mass terms in (\ref{fermion3}) and
the twisted boundary conditions caused by $\Omega(z)$ in (\ref{gaugeT1}).
The $z$-dependence of the solutions to (\ref{Feq1}) is expressed
in terms of the Bessel functions.  The basis functions are given by
\beqn
&&\hskip -1cm
\begin{pmatrix} C_L \cr S_L \end{pmatrix} (z; \lambda, c) 
= \pm  \frac{\pi}{2} \lambda \sqrt{z z_L}
        \, F_{c+\onehalf, c \mp \onehalf} (\lambda z, \lambda z_L) ~, \cr
\noalign{\kern 10pt}
&&\hskip -1cm
\begin{pmatrix} C_R \cr S_R \end{pmatrix} (z; \lambda, c) 
 = \mp  \frac{\pi}{2} \lambda \sqrt{z z_L}
        \, F_{c-\onehalf, c \pm \onehalf} (\lambda z, \lambda z_L) ~, 
\label{Bessel1}
\eeqn
where $F_{\alpha,\beta}(u,v)=J_{\alpha}(u)Y_{\beta}(v)-Y_{\alpha}(u)J_{\beta}(v)$.
They satisfy the relations
$S_L(z; \lambda, -c) = -S_R (z; \lambda, c)$ and 
$C_L C_R - S_L S_R = 1$.  They also obey  the boundary conditions that
$C_R= C_L = 1$,  $D_- C_R = D_+ C_L =0$, 
$S_R = S_L = 0 $ and  $D_- S_R = D_+ S_L =\lambda$ at  $z= z_L$.
Further $D_\pm$ links them by
$D_+(C_L, S_L) = \lambda (S_R, C_R)$ and
$D_-(C_R, S_R) = \lambda (S_L, C_L)$.

In the $Q_\EM = \frac{2}{3}$ sector (the top sector) $U$, $B$, $t$, $t'$, 
$\hat U_R$ and $\hat B_R$ mix with each other. 
The top quark component $t(x)$ in four dimensions is contained 
in these fields in the form
\beqn
&&\hskip -1cm
\begin{pmatrix} \tilde U_L \cr 
     (\tilde B_L \pm \tilde t_L)/\sqrt{2} \cr
     \tilde t'_L \end{pmatrix} (x,z) = \sqrt{k}
\begin{pmatrix}a_U C_L(z; \lambda, c_2) \cr 
    a_{B \pm t} C_L(z; \lambda, c_1) \cr 
    a_{t'} S_L(z; \lambda, c_1)  \end{pmatrix} \, t_L(x) \cr
\noalign{\kern 10pt}
&&\hskip -1cm
\begin{pmatrix} \tilde U_R \cr 
     (\tilde B_R \pm \tilde t_R) /\sqrt{2} \cr
     \tilde t'_R \end{pmatrix} (x,z) = \sqrt{k}
\begin{pmatrix}a_U S_R(z; \lambda, c_2) \cr 
    a_{B \pm t} S_R(z; \lambda, c_1) \cr 
    a_{t'} C_R(z; \lambda, c_1)  \end{pmatrix} \, t_R(x)~.
\label{top1}
\eeqn
The brane fermions are related to the bulk fermions by
\beeq
\hat U_R(x) = \frac{2}{\mu_2^*}  U_R (x, 1^+) 
      = \frac{2}{\tilde \mu^*}  t_R (x, 1^+) ~~,~~
\hat B_R(x) = \frac{2}{\mu_1^*}  B_R (x, 1^+) 
\label{top2}
\eneq
as follows from the equations of motion.  We note that 
$U_R$, $t_R$  and $B_R$ develop discontinuities at the Planck brane.
The top quark mass is given by $m_t = k \lambda$.  
The coefficients $a_j$'s are common to both left- and right-handed
components as a consequence of  the equations of motion in the bulk
($\bar \sigma \dd \tilde U_R = k D_+ \tilde U_L$ etc.) with the 
normalization $\bar \sigma \dd \,  t_R(x) = m_t \, t_L(x)$.

\begin{figure}[t,b]
\centering  \leavevmode
\includegraphics[height=5cm]{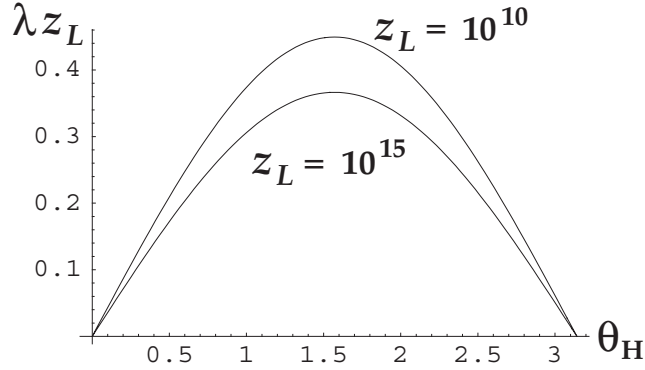}
\caption{The  $\theta_H$-dependence of $\lambda  z_L$ of the top quark
for $z_L = 10^{10}$ and $z_L = 10^{15}$. The top mass is given by $m_t = \lambda k$. 
The plots fit well with $\kappa \sin \theta_H$ as in (\ref{spectrum2}).
}
\label{top-fig}
\end{figure}

The eigenvalue $\lambda$ and coefficients $a_j$'s are
determined from the boundary conditions.  
The details of the computations were given in ref.\ \cite{HOOS}.  
Let us denote $s_H = \sin \theta_H$, $c_H = \cos \theta_H$,  and
$C_L^{(j)} = C_L(1; \lambda, c_j)$ etc.
The coefficients satisfy 
$s_H a_{B-t}C_L^{(1)} =  c_H a_{t'}S_L^{(1)}$ and
\beqn
&&\hskip -1cm
K ~ \left[
\begin{array}{ccc}a_U\\
 \big( a_{B+t}- c_H^{-1} a_{B-t} \big) / \sqrt{2}\\
 \big( a_{B+t}+c_H^{-1} a_{B-t} \big) / \sqrt{2}
\end{array} \right]=0 ~~, \cr
\noalign{\kern 10pt}
&&\hskip -1cm
K = 
\left[\begin{array}{ccc}
\lambda S_R^{(2)}-\myfrac{|\mu_2|^2}{2k} C_L^{(2)} 
& -\myfrac{\mu_2^* \tilde{\mu}}{2k}C_L^{(1)} & 0\\
-\myfrac{\tilde\mu^* \mu_2}{2k}C_L^{(2)}
&\lambda \bar S^{(1)}  -\myfrac{|\tilde \mu|^2}{2k}  C_L^{(1)} 
& - \myfrac{\lambda}{2} \myfrac{s_H^2}{S_L^{(1)}}\\
0 & - \lambda \myfrac{s_H^2}{S_L^{(1)}} 
& 2 \lambda \bar S^{(1)} - \myfrac{|\mu_1|^2}{k}  C_L^{(1)}
\end{array}\right] 
\label{top3}
\eeqn
where $\bar S^{(1)} = S_R^{(1)}+(s_H^2/ 2S_L^{(1)})$.
The top mass, or the eigenvalue $\lambda$, is determined by the condition 
${\rm det} \, K = 0$.  When $|\mu_j|^2, |\tilde \mu|^2 \gg m_\KK$, 
the equation is approximated,  to high accuracy,  by
\beeq
|\mu_2|^2 C_L^{(2)}
\bigg\{S_R^{(1)}+\frac{s_H^2}{2S_L^{(1)}} \bigg\}
+|\tilde{\mu}|^2 C_L^{(1)}S_R^{(2)}=0 ~.
\label{spectrum1}
\eneq
The first term in (\ref{spectrum1}) dominates
over the second.  With given $z_L$,    $c_1$ is fixed so as to reproduce 
the observed $m_t \sim 172 \,$GeV at $\theta_H = \onehalf \pi$.
See  Table \ref{top-table}.
With these parameters fixed, the $\theta_H$-dependence of $m_t$
is determined numerically, which is depicted in Fig.\ \ref{top-fig} for $z_L = 10^{10}$
and $10^{15}$.  The curves fit well with
\beeq
m_t\sim\frac{m_{KK}}{\sqrt2\pi}\sqrt{1-4c_1^2} ~ |\sin\theta_H|
\label{spectrum2}
\eneq
with an error of $2.0\%\sim4.0\%$.  
The top mass $m_t = \lambda k$  vanishes at $\theta_H=0$ as the chiral 
symmetry is restored.
The effective potential $V_\eff (\theta_H)$ is evaluated
from the $\theta_H$-dependence of the mass spectrum.  It was found that
the contribution from the top quark dominates over those from gauge fields
and other fermions.  $V_\eff$ is minimized at $\theta_H = \pm \onehalf \pi$.

\begin{table}[b,t]
\begin{center}
\begin{tabular}{|c||c|c|c|c|}
\hline
$z_L=e^{kL}$& $k$(GeV) & $\lambda(\theta_H=\pi/2)$ & c & $m_{KK}$(TeV)\\
\hline\hline
$10^{15}$ & $4.70\times 10^{17}$ & $3.66\times 10^{-16}$ & $0.432$ & 1.48\\
\hline
$10^{10}$ & $3.83\times 10^{12}$ & $4.49\times 10^{-11}$ & $0.396$ & 1.20\\
\hline
\end{tabular}
\end{center}
\caption{With the value of $z_L$ given, $k$, $\lambda$, $c_1=c_2=c$ are determined. 
Input parameters are the $W$ boson mass $m_W$=80.40 GeV and 
the top quark mass $m_t$=172 GeV}
\label{top-table}
\end{table}

To be definite, let us take  $\mu_j, \tilde\mu >0$
given by 
\beeq
\mu_1^2 = \mu_2^2 = 10^{10} \, {\rm GeV} ~,~
\tilde \mu^2 = 5.96 \times 10^6 \, {\rm GeV} ~, 
\label{massvalue1}
\eneq
which, a posteriori, leads to  the value $m_b/m_t \sim 4.2/172$ for $c_1=c_2$.  
With the value $\lambda$ for the top quark, 
$\lambda S_R/[(\mu_2^2/2k)C_L]$ in the matrix $K$ in (\ref{top3}),
for instance, is  $O(10^{-15})$ so that the equation  (\ref{top3}) 
is well approximated by 
\beeq
\begin{pmatrix} |\mu_2|^2 C_L^{(2)} & \mu_2^* \tilde \mu C_L^{(1)} & 0 \cr
\tilde \mu^* \mu_2 C_L^{(2)} & |\tilde\mu |^2 C_L^{(1)}  & 0 \cr
0 & 0 & 2 |\mu_1|^2 C_L^{(1)}   \end{pmatrix} 
 \left[
\begin{array}{ccc}a_U\\
 \big( a_{B+t}- c_H^{-1} a_{B-t} \big) / \sqrt{2}\\
 \big( a_{B+t}+c_H^{-1} a_{B-t} \big) / \sqrt{2}
\end{array} \right] \sim 0 ~.
\label{top4}
\eneq
It follows that
\beeq
\big[ ~ a_{B-t}, \, a_U, \, a_{t'} ~ \big] \sim
\Bigg[ -  c_H , -\myfrac{\sqrt{2} \tilde{\mu}  C_L^{(1)}}{\mu_2  C_L^{(2)}} , 
- \myfrac{s_H C_L^{(1)}}{S_L^{(1)} } ~ \Bigg]
~ a_{B+t} ~.
\label{top5}
\eneq
The coefficient  $a_{B+t}$ is determined so as to have canonical normalization for the 
kinetic term of $t_L(x)$.  Note that $\lambda$ depends on $\theta_H$.

In the $Q_\EM = - \frac{1}{3}$ sector (the bottom sector) 
$b$,  $D$,  $X$,  $b'$, $\hat D_R$ and $\hat X_R$ mix with each other.  As in 
the top sector,  the bottom quark component $b(x)$ in four
dimensions appears as
\beqn
&&\hskip -1cm
\begin{pmatrix} \tilde b_L \cr 
     (\tilde D_L \pm \tilde X_L)/\sqrt{2} \cr
     \tilde b'_L \end{pmatrix} (x,z) = \sqrt{k} 
\begin{pmatrix}a_b C_L(z; \lambda, c_1) \cr 
    a_{D \pm X} C_L(z; \lambda, c_2) \cr 
    a_{b'} S_L(z; \lambda, c_2)  \end{pmatrix} \, b_L(x) \cr
\noalign{\kern 10pt}
&&\hskip -1cm
\begin{pmatrix} \tilde b_R \cr 
     (\tilde D_R \pm \tilde X_R) /\sqrt{2} \cr
     \tilde b'_R \end{pmatrix} (x,z) = \sqrt{k}
\begin{pmatrix}a_b S_R(z; \lambda, c_1) \cr 
    a_{D \pm X} S_R(z; \lambda, c_2) \cr 
    a_{b'} C_R(z; \lambda, c_2)  \end{pmatrix} \, b_R(x)~.
\label{bottom1}
\eeqn
The brane fermions are related to the bulk fermions by
\beeq
\hat D_R(x) = \frac{2}{\mu_2^*}  D_R (x, 1^+) 
      = \frac{2}{\tilde \mu^*}  b_R (x, 1^+) ~~,~~
\hat X_R(x) = \frac{2}{\mu_3^*}  X_R (x, 1^+) 
\label{bottom2}
\eneq
The equation corresponding to (\ref{top3}) is obtained by replacing
$(U,B,t)$ by $(b,D,X)$ and interchanging $(c_1, c_2)$, $(\mu_1, \mu_3)$
and $(\mu_2, \tilde \mu)$.
In the same approximation as in the top case the bottom mass and
the coefficients $a_j$'s are found, for $0< c_1, c_2 < \onehalf$, to be
\beeq
m_b \sim\sqrt{ \frac{1 + 2c_2}{1+2c_1} }  \Big| \frac{\tilde \mu}{\mu_2} \Big| 
z_L^{c_1 - c_2} ~ m_t
\label{bottommass1}
\eneq 
and
\beeq
\big[ ~ a_{D+X}, \, a_{D-X}, \, a_{b'} ~ \big] \sim
\Bigg[- 1 ,   c_H  , 
 \myfrac{s_H C_L^{(2)}}{S_L^{(2)} } ~ \Bigg]
~\frac{\tilde \mu C_L^{(1)}}{\sqrt{2}\mu_2 C_L^{(2)}} ~ a_{b} ~.
\label{bottom3}
\eneq

With the wave functions of the top and bottom quarks at hand, 
one can  evaluate their Yukawa couplings  in two manners.  
In the Kaluza-Klein approach we insert the wave functions into the 
five-dimensional Lagrangian density 
${\cal L}_{\rm bulk}^{\rm fermion}  + {\cal L}^{\rm brane}_{\rm mass}$
and integrate
over the fifth dimensional coordinate to obtain four-dimensional Lagrangian. 
The part
$k^{-1} \sum_{j=1}^2 \overline{\tilde \Psi}_j (\gamma \dd)_{d=4} \tilde \Psi_j$
gives the four-dimensional kinetic terms for the top and bottom quarks.
The part with the covariant derivative in the fifth coordinate
\beeq
\sum_{j=1}^2  \Big\{ 
-i {\tilde\Psi}_{jL}^\dagger \big( D_-(c_j) + ig_A \tilde A_z \big) {\tilde\Psi}_{jR}
+ i {\tilde\Psi}_{jR}^\dagger \big( D_+(c_j) - ig_A \tilde A_z \big) 
      {\tilde\Psi}_{jL} \Big\}
\label{Lag1}
\eneq
generates both the masses and Yukawa couplings of the top and bottom quarks.
The 4D Higgs field is contained in the  gauge potential $A_z$.
The vev $v$ of $\phi^4(x)$  in (\ref{Higgs1}) is related to 
$\theta_H$ by (\ref{Higgs2})  
and its  fluctuation around $v$ corresponds to the neutral Higgs field $H(x)$.  
Hence the relevant part of the gauge potential is expressed as
\beeq
A_z (x, z) = \hat \theta_H (x) \cdot \frac{2\sqrt{2} \,  z}{z_L^2 -1} 
\cdot T^{\hat 4} + \cdots 
\label{Az1}
\eneq
in the original gauge where
\beeq
\hat \theta_H (x) = \theta_H + \frac{H(x)}{f_H}
~~,~~
f_H = \frac{2}{g_A} \sqrt{\frac{k}{z_L^2 -1}}
\sim \frac{2}{\sqrt{kL}} \frac{m_\KK}{\pi g} ~~. 
\label{effective1}
\eneq
In the twisted gauge defined in (\ref{gaugeT1}), 
$\tilde A_z^c = \la \tilde A_z \ra$ vanishes, 
$\tilde A_z(x,z)$ being expanded as in (\ref{Az1}) with $\hat \theta_H$
replaced by $H(x)/f_H$.

The Yukawa coupling originates from 
$g_A( {\Psi}_{L}^\dagger A_z   {\Psi}_{R} +  {\Psi}_{R}^\dagger  A_z   {\Psi}_{L})$
or $g_A( {\tilde\Psi}_{L}^\dagger \tilde A_z   {\tilde\Psi}_{R}
+  {\tilde\Psi}_{R}^\dagger \tilde A_z   {\tilde\Psi}_{L})$, whereas
the mass term comes from 
$-i {\Psi}_{L}^\dagger \big( D_- + ig_A  A_z^c \big) {\Psi}_{R}
+ i {\Psi}_{R}^\dagger \big( D_+ - ig_A  A_z^c \big)  {\Psi}_{L}$
in the original gauge 
or $ -i {\tilde\Psi}_{jL}^\dagger D_- {\tilde\Psi}_{jR}
+ i {\tilde\Psi}_{jR}^\dagger  D_+  {\tilde\Psi}_{jL}$ in the twisted gauge.  
The terms involving $D_\pm$ are  important.
With the wave function in (\ref{Higgs1}), (\ref{top1}) and (\ref{bottom1}) inserted, 
$\vphi_H (z) {\tilde\Psi}_{jL}^\dagger T^{\hat 4} \tilde \Psi_{jR}$ 
($\vphi_H (z) {\tilde\Psi}_{jR}^\dagger  T^{\hat 4} \tilde \Psi_{jL}$)
has different $z$-dependence 
from ${\tilde\Psi}_{jL}^\dagger D_- \tilde \Psi_{jR}$ 
(${\tilde\Psi}_{jR}^\dagger D_+ \tilde \Psi_{jL}$).    
After the integration over $z$,  
the Yukawa coupling is not proportional to the fermion mass in the RS spacetime.  
We also recall that  a large gauge transformation generates 
$\theta_H \go \theta_H + 2 \pi$ so that
the mass spectrum remains invariant under the shift 
$\theta_H \go \theta_H + 2 \pi$, or equivalently 
under $H(x) \go H(x) + 2\pi f_H$.  The mass is a periodic, nonlinear function 
of $\theta_H$.  (There is no level-crossing in the RS spacetime.) 
The  nonlinearity in the relation between the Yukawa coupling and mass 
is confirmed by direct evaluation described below.

Let us define the normalized coefficients $a'^{L,R}_j$   by 
\beqn
&&\hskip -1cm 
\big(a'^L_U , a'^L_{B\pm t}, a'^L_{t'} \big) = 
\big( \sqrt{N_{C_L}^{(2)}  } \, a_U , \sqrt{N_{C_L}^{(1)} } \, a_{B \pm t} , 
\sqrt{N_{S_L}^{(1)} } \, a_{t'}  \big) ~, \cr
\noalign{\kern 10pt}
&&\hskip -1cm
\big(a'^R_U , a'^R_{B\pm t}, a'^R_{t'} \big) = 
\big( \sqrt{N_{S_R}^{(2)} } \, a_U , \sqrt{N_{S_R}^{(1)} } \, a_{B \pm t} , 
\sqrt{N_{C_R}^{(1)} } \, a_{t'}  \big)  ~, 
\label{normalization1}
\eeqn
where $N_{C_L}^{(j)}  = \int_1^{z_L} dz \,  C_L (z; \lambda, c_j)^2$ etc..
Then the free part of the Lagrangian for the top quark is found to be
\beqn
&&\hskip -1cm
{\cal L}^{4D}_{{\rm free}} \sim 
- P_L  it^{\dagger}_L \sigma\dd  t_L +P_R  it^{\dagger}_R \sigma\dd t_R 
+ \lambda k  \frac{P_L + P_R}{2} 
(  it^{\dagger}_L t_R - i t^{\dagger}_R t_L) ~, \cr
\noalign{\kern 5pt}
&&\hskip -1cm
P_{L,R} =  |a'^{L,R}_U|^2 +   |a'^{L,R}_{B+t} |^2  +  |a'^{L,R}_{B-t} |^2  
+ |a'^{L,R}_{t'}|^2 ~.
\label{4Dtop1}
\eeqn
The contributions coming from the brane mass term 
${\cal L}^{\rm brane}_{\rm mass}$   turn  out 
$O(10^{-15})$ smaller than $P_L$ and $P_R$, and can be ignored.   

Recall that $D_- S_R = \lambda C_L$
and $D_+ C_L = \lambda S_R$, from which it follows that
$N_{C_L} = N_{S_R} + \lambda^{-1} S_R C_L |_{z=1}$.  Hence
\beqn
&&\hskip -1cm
P_L = P_R + \frac{1}{\lambda} \Big\{ S_R^{(2)} C_L^{(2)} |a_U|^2
+ S_R^{(1)} C_L^{(1)} \big( |a_{B+t}|^2 + |a_{B-t}|^2 \big) + 
S_L^{(1)} C_R^{(1)} |a_{t'}|^2 \Big\}  \cr
\noalign{\kern 10pt}
&&\hskip -0.4cm
= P_R + \frac{2}{\lambda} |a_{B+t}|^2 C_L^{(1)}
\bigg\{ S_R^{(1)} + \frac{s_H^2}{2 S_L^{(1)}} 
+ \frac{|\tilde\mu|^2}{|\mu_2|^2} \frac{S_R^{(2)}C_L^{(1)}}{C_L^{(2)}} \bigg\} \cr
\noalign{\kern 10pt}
&&\hskip -0.4cm
= P_R ~.
\label{identity1}
\eeqn
The relations (\ref{top5}) and $C_L C_R - S_L S_R =1$ have been used 
in the second equality.  
The last equality follows from  the relation  (\ref{spectrum1}) determining 
the mass spectrum.
Let us adopt the normalization  $P_L = P_R =1$  with which  
the top mass appears as $\lambda k$ in  (\ref{4Dtop1})  as it should.
The coefficients $a'^L_j$ and $a'^R_j$ represent how much portion of each field contains the left- and right-handed top  quark, respectively.  

Similarly the normalized coefficients 
$a'^{L,R}_b$,  $a'^{L,R}_{D\pm X}$,  $a'^{L,R}_{b'}$ are determined.
The numerical values are tabulated in Table \ref{norm-table}.  
The numerical values for the dominant terms 
($a'^{L}_{B\pm t}$,  $a'^{L,R}_{t'}$, $a'^{L}_b$,  $a'^{L}_{D\pm X}$,  
and $a'^{R}_{b'}$) do not vary very much with $z_L$ in the range $10^{10}$
to $10^{15}$.
In the $\theta_H = 0$ limit, the four-dimensional $t_L(x)$ and $t_R(x)$
are mostly contained in the five-dimensional $t$ and $t'$, respectively.
At  $\theta_H = \onehalf \pi$,    $t_L(x)$ resides
in the $(B+t)/\sqrt{2}$ and $t'$ components,  whereas 
  $t_R(x)$ remains in $t'$.
The four-dimensional $b_L(x)$ and $b_R(x)$ are mostly contained,
for any value of $\theta_H$,  
in the five-dimensional $b$ and $b'$, respectively.

\begin{table}[b,t]
\begin{center}
\begin{tabular}{|ccc|rrr|rrr|} 
\noalign{\kern 15pt}
\hline
\multicolumn{3}{|c|}{~} & \multicolumn{3}{|c|}{$\theta_H = 0$} 
   & \multicolumn{3}{|c|}{$\theta_H = \onehalf \pi$}    \\ \hline \hline 
$a_U$ & $a'^L_U$ & $a'^R_U$
& $2.9 \times 10^{-10}$ & $ 0.024$ & $5.1\times 10^{-5}$  
& $3.0 \times 10^{-10}$ & $ 0.025$ & $ 0.0017$   \\ \hline
$a_{B+t}$ & $a'^L_{B+t}$ & $a'^R_{B+t}$
& $1.2 \times 10^{-8}$ & $ 0.71$ & $0.0015$  
& $1.2 \times 10^{-8}$ & $0.73$ & $0.050$ \\ \hline
$a_{B-t}$ & $a'^L_{B-t}$ & $a'^R_{B-t}$
& $- 1.2 \times 10^{-8}$ & $ -0.71$ & $ -0.0015$  & $0$ & $0$ & $0$ \\ \hline
$a_{t'}$ & $a'^L_{t'}$ & $a'^R_{t'}$
& $4.3 \times 10^{-8}$ & $ 0.021$ & $1.0$  & $4.4 \times 10^{-8}$ & $ 0.69$ & $1.0$ \\ \hline \hline
$a_{b}$ & $a'^L_{b}$ & $a'^R_{b}$
& $1.2 \times 10^{-8}$ & $1.0$ & $ 5.1\times 10^{-5}$
& $1.2 \times 10^{-8}$ & $1.0$ & $0.0016$   \\ \hline
$a_{D+X}$ & $ a'^L_{D+X}$ & $a'^R_{D+X}$
& $2.9 \times 10^{-10}$ & $0.017$ & $8.8\times 10^{-7}$
& $2.9 \times 10^{-10}$ & $ 0.017$ & $ 2.8\times 10^{-5}$ \\ \hline
$a_{D-X}$ & $ a'^L_{D-X}$ & $a'^R_{D-X}$
& $-2.9 \times 10^{-10}$ & $ -0.017$ & $ -8.8\times 10^{-7}$ & $0$~ & $0$~ & $0$~ \\ \hline
$a_{b'}$ & $a'^L_{b'}$ & $a'^R_{b'}$
& $4.3 \times 10^{-8}$~ & $ 0.00051$ & $ 1.0$  & $4.3 \times 10^{-8}$~ & $ 0.016$ & $1.0$ \\ \hline
\end{tabular}
\end{center}
\caption{The coefficients (\ref{normalization1}) of the wave functions 
of the top and bottom quarks at $\theta_H = 0$ and $\onehalf \pi$, 
evaluated for $c_1=c_2=0.43$, $z_L=10^{15}$, and $\mu_j, \tilde \mu$ in 
(\ref{massvalue1}).}
\label{norm-table}
\end{table}

\begin{figure}[t,b]
\centering  \leavevmode
\includegraphics[height=5cm]{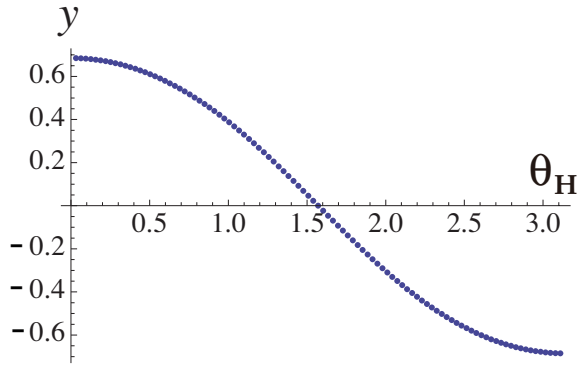}
\caption{The  $\theta_H$-dependence of the Yukawa coupling for the top 
quark for $z_L=10^{15}$.  The curve is well approximated by a cosine
curve.  The curve has little dependence on $z_L$.}
\label{yukawa1-fig}
\end{figure}

The Yukawa couplings are evaluated in the same manner.
\ignore{
emerge from the couplings of fermions to 
the extra-dimensional component $A_z$ in 
$- ig_A\sum_a  \bar \Psi_a  \Gamma^5 {e_5}^z A_z \Psi_a \equiv {\cal L}_Y$.
The vev $v$ of $\phi^4(x)$ in (\ref{Higgs1}) is related to $\theta_H$ by (\ref{Higgs2})  
and its  fluctuation around $v$ corresponds to the neutral Higgs field $H(x)$.  
In the twisted gauge the vev of $\tilde A^{\hat 4}_z$ vanishes.
}
Inserting $\tilde A^{\hat 4}_z = H(x) \vphi_H(z)$ and the wave functions (\ref{top1})
into (\ref{Lag1}) in the twisted gauge, one finds, for the top quark, 
\beqn
&&\hskip -1cm
\sqrt{\det g}~  {\cal L}_Y= -\frac{i}{2} g_A H \vphi_H(z)
\Big\{  \tilde{t'}_R^{\dagger} (\tilde{t}_L-\tilde{B}_L)
+ \tilde{t'}_L^{\dagger} (\tilde{t}_R-\tilde{B}_R) - ({\rm h.c.)} \Big\}  \cr
\noalign{\kern 5pt}
&&\hskip +1.cm
= -\frac{i}{\sqrt{2}} g_A k  ~ a_{t'} a_{B-t} ~
\vphi_H(z)  H(x)  \big\{ t^\dagger_R t_L(x)  - t^\dagger_L t_R(x) \big\} ~.
\label{yukawa1}
\eeqn
The overall phase of the $a_j$'s  has been taken to be real.
By making use of (\ref{top5}) and integrating over $z$,  the 4D
Yukawa coupling constant  in
${\cal L}^{\rm 4D}_{\rm Yukawa} = iy H (t_L^\dagger t_R - t_R^\dagger t_L)$ 
is found to be
\beqn
&&\hskip -1cm
y (\theta_H)  =   \frac{g \sqrt{kL ( z_L^2 -1) } \, 
 s_H c_H C_L^{(1)} }{4 S_L^{(1)} \bar P} ~, \cr
\noalign{\kern 10pt}
&&\hskip -1cm
\bar P = \frac{1 + c_H^2}{2} N_{C_L}^{(1)} 
+ \frac{s_H^2}{2} \bigg( \frac{ C_L^{(1)} }{S_L^{(1)}} \bigg)^2 N_{S_L}^{(1)}
+ \frac{|\tilde \mu|^2}{|\mu_2|^2} 
     \bigg( \frac{ C_L^{(1)} }{C_L^{(2)} } \bigg)^2  N_{C_L}^{(2)} ~.
\label{yukawa2}
\eeqn
Note that $s_H/N_{S_L}^{(1)}$ remains finite  in the  $s_H \go 0$ limit.
The $\theta_H$-dependence of $y (\theta_H)$ for the top quark is depicted in
fig.\  \ref{yukawa1-fig}, which is well approximated by the cosine curve.
It is seen that $y$ vanishes at $\theta_H = \onehalf \pi$.  
The result  for the bottom quark is similar to that for the top quark, with
a magnitude scaled down by a factor $m_b/m_t$. 

So far we have evaluated the masses and Yukawa couplings of the top and
bottom quarks in the Kaluza-Klein approach.  One can develop an effective
interaction approach \cite{Panico2, Sakamura1, Giudice1}
to concisely summarize the  results.   It enables
us for deducing the Higgs couplings in higher order as well.

In the original gauge $\theta_H$ and $H(x)$  always appear in the
combination $\hat \theta_H (x)$ in (\ref{effective1}).  
Therefore the effective local interactions at low energies, which manifest
significant deviation from the standard model, can be  written in the form
\beqn
&&\hskip -1cm
{\cal L}_\eff  = - V_\eff (\hat \theta_H) - m_W(\hat \theta_H)^2 W^\dagger_\mu W^\mu
- \onehalf m_Z(\hat \theta_H)^2 Z_\mu Z^\mu \cr
\noalign{\kern 5pt}
&&\hskip 2.cm
- \sum_f m_f(\hat \theta_H) \psibar_f \psi_f ~.
\label{effective2}
\eeqn
The key feature is that $ \theta_H$ is a phase variable so that
${\cal L}_\eff$ is periodic in $\hat \theta_H$ with a period $2\pi$.
The first term is the effective potential for $\hat \theta_H$.  As shown in 
ref.\  \cite{YH1}, $V_\eff$ is finite and the value of $\theta_H$ is unambiguously 
determined  by the location of its global minimum.  The Higgs mass $m_H$, given by
$m_H^2 = V_\eff^{(2)} (\theta_H)/f_H^2$, is predicted to be finite.   
$m_W(\hat \theta_H)$ and $m_Z(\hat \theta_H)$ in the $SO(5) \times U(1)_X$ 
model in the RS spacetime has been evaluated in refs.\ \cite{SH1, HS2};
\beeq
m_W(\hat \theta_H) \sim \cos \theta_W \, m_Z(\hat \theta_H)
\sim \onehalf g f_H \sin \hat \theta_H 
\label{effective3}
\eneq
where $m_W = m_W(\theta_H)$, $m_Z = m_Z(\theta_H)$, and $\theta_W$ 
is the Weinberg angle.   Expanding $m_W(\hat \theta_H)^2$ and 
$m_Z(\hat \theta_H)^2$  in (\ref{effective2}) in a power series in $H$, one finds
that $WWH$ and $ZZH$ couplings are suppressed by a factor
$\cos \theta_H$ compared with those in the standard model.
For the $WWHH$ and $ZZHH$ couplings the suppression factor becomes
$\cos 2 \theta_H$.  As demonstrated by Sakamura,  it includes the contributions of 
the KK towers of $W$ and $Z$ in the intermediate states.\cite{Sakamura1}
The effective interactions contain contributions coming from heavy KK excited states.

We apply the same argument to the last term in (\ref{effective2}). 
In this approach the Yukawa coupling $y_f H \psibar_f \psi_f$ is related to 
the mass  by
\beeq
y_f(\theta_H) = \frac{1}{f_H}  \frac{d m_f (\theta_H) }{d\theta_H} ~.
\label{yukawa3}
\eneq
The top quark mass $m_t(\theta_H)$ is  determined from 
(\ref{spectrum1}) as a function of $\theta_H$.  
Its derivative $d m_t(\theta_H) / d \theta_H$
is compared with the Yukawa coupling $y_t(\theta_H)$ in (\ref{yukawa2})
determined in the Kaluza-Klein approach.  
We have numerically confirmed  that the equality (\ref{yukawa3}) between the two 
holds with an error less than  $0.3\,$\% in the entire region of $\theta_H$, 
which establishes the validity and usefulness of the effective interaction approach.
As is seen in fig.\  \ref{top-fig}, the mass $m_t(\theta_H)$ reaches the maximum
at $\theta_H = \onehalf \pi$.  
The relation (\ref{yukawa3}) implies that the Yukawa coupling 
$y_t(\theta_H)$ vanishes there, which,   independently, is shown in 
the Kaluza-Klein approach as well.
In the effective interaction approach the $HH\psibar_f \psi_f$ coupling, 
is given by $m_f^{(2)} (\theta_H)/f_H^2$.   
In the HOOS model $m_f (\hat \theta_H) \sim \kappa_f \sin \hat \theta_H$
and $\theta_H = \onehalf \pi$.
Although the Yukawa coupling $y_f$ vanishes, the $HH\psibar_f \psi_f$
coupling is nonvanishing ($\sim - m_f/f_H^2$).
The KK excited states of $\psi_f$ contribute in the intermediate states
for the $HH\psibar_f \psi_f$ coupling.  

In this paper we have given  detailed analysis of the Yukawa couplings in the 
$SO(5) \times U(1)$ gauge-Higgs unification model, particularly in the 
HOOS model\cite{HOOS}.  We have determined
the wave functions of the top and bottom quarks in the extra-dimensional
space, with which the Yukawa couplings are evaluated numerically in the
Kaluza-Klein approach.
We have also shown that all the results are concisely cast in the form of  
the effective interactions.

The phenomenological implication  is significant.  In the gauge-Higgs unification 
scenario the large deviation from the standard model of electroweak 
interactions appears in the Higgs couplings.  All of the $WWH$, $ZZH$,
and Yukawa couplings are suppressed by a factor $\cos \theta_H$,
which can be checked in the forthcoming experiments at LHC.
In the HOOS model, in particular, $\theta_H = \onehalf \pi$ is dynamically
realized, leading to completely new phenomenology.  
The Higgs particle becomes stable  in the low energy effective theory at the tree level.   
It is interesting to see whether or not the Higgs particle can decay 
at all through heavy KK excited states.  
We will come back on this issue in a separate paper in more detail.

\vskip 1cm

\leftline{\bf Acknowledgments}
This work was supported in part 
by  Scientific Grants from the Ministry of Education and Science, 
Grant No.\ 20244028, Grant No.\ 20025004,  and Grant No.\ 50324744 (Y.H.).

\vskip 1cm

\def\jnl#1#2#3#4{{#1}{\bf #2} (#4) #3}

\def\Zphys{{\em Z.\ Phys.} }
\def\jssc{{\em J.\ Solid State Chem.\ }}
\def\jpsJ{{\em J.\ Phys.\ Soc.\ Japan }}
\def\ptps{{\em Prog.\ Theoret.\ Phys.\ Suppl.\ }}
\def\PTP{{\em Prog.\ Theoret.\ Phys.\  }}

\def\JMP{{\em J. Math.\ Phys.} }
\def\NPB{{\em Nucl.\ Phys.} B}
\def\NP{{\em Nucl.\ Phys.} }
\def\PLB{{\em Phys.\ Lett.} B}
\def\PL{{\em Phys.\ Lett.} }
\def\PRL{\em Phys.\ Rev.\ Lett. }
\def\PRB{{\em Phys.\ Rev.} B}
\def\PRD{{\em Phys.\ Rev.} D}
\def\PRe{{\em Phys.\ Rep.} }
\def\AP{{\em Ann.\ Phys.\ (N.Y.)} }
\def\RMP{{\em Rev.\ Mod.\ Phys.} }
\def\ZPC{{\em Z.\ Phys.} C}
\def\SCI{\em Science}
\def\CMP{\em Comm.\ Math.\ Phys. }
\def\MPLA{{\em Mod.\ Phys.\ Lett.} A}
\def\IJMPA{{\em Int.\ J.\ Mod.\ Phys.} A}
\def\IJMPB{{\em Int.\ J.\ Mod.\ Phys.} B}
\def\EPJC{{\em Eur.\ Phys.\ J.} C}
\def\PR{{\em Phys.\ Rev.} }
\def\JHEP{{\em JHEP} }
\def\cmp{{\em Com.\ Math.\ Phys.}}
\def\JPA{{\em J.\  Phys.} A}
\def\JPG{{\em J.\  Phys.} G}
\def\NJP{{\em New.\ J.\  Phys.} }
\def\CQG{\em Class.\ Quant.\ Grav. }
\def\ATMP{{\em Adv.\ Theoret.\ Math.\ Phys.} }
\def\ibid{{\em ibid.} }

\renewenvironment{thebibliography}[1]
         {\begin{list}{[$\,$\arabic{enumi}$\,$]}  
         {\usecounter{enumi}\setlength{\parsep}{0pt}
          \setlength{\itemsep}{0pt}  \renewcommand{\baselinestretch}{1.2}
          \settowidth
         {\labelwidth}{#1 ~ ~}\sloppy}}{\end{list}}

\def\reftitle#1{}                

\end{document}